\documentclass{elsart3p} 

\usepackage{graphicx}

\usepackage{amssymb}

\def\dbar{\overline{D}{}^{\,0}}
\def\cp{$CP$}

\begin{document}

\begin{frontmatter}

  \title{Measurement of $D^0$-$\dbar$ mixing and search for $CP$ violation
    in $D^0 \to K^+K^-, \pi^+\pi^-$ decays with the full Belle data set}


  \collab{Belle Collaboration}

  \author[JSI]{M.~Stari\v{c}} 
  \author[Tabuk]{A.~Abdesselam} 
  \author[KEK,Sokendai]{I.~Adachi} 
  \author[Tokyo]{H.~Aihara} 
  \author[BINP,NSU]{K.~Arinstein} 
  \author[PNNL]{D.~M.~Asner} 
  \author[MIPT,ITEP]{T.~Aushev} 
  \author[Tabuk]{R.~Ayad} 
  \author[Tata]{T.~Aziz} 
  \author[Tata]{V.~Babu} 
% ...

% ...
\address[BINP]{Budker Institute of Nuclear Physics SB RAS, Novosibirsk 630090, Russian Federation}
% ...
\address[Sokendai]{SOKENDAI (The Graduate University for Advanced Studies), Hayama 240-0193, Japan}
% ...
\address[KEK]{High Energy Accelerator Research Organization (KEK), Tsukuba 305-0801, Japan}
% ...
\address[ITEP]{Institute for Theoretical and Experimental Physics, Moscow 117218, Russian Federation}
\address[JSI]{J. Stefan Institute, 1000 Ljubljana, Slovenia}
% ...
\address[MIPT]{Moscow Institute of Physics and Technology, Moscow Region 141700, Russian Federation}
% ...
\address[NSU]{Novosibirsk State University, Novosibirsk 630090, Russian Federation}
% ...
\address[PNNL]{Pacific Northwest National Laboratory, Richland, WA 99352, USA}
% ...
\address[Tabuk]{Department of Physics, Faculty of Science, University of Tabuk, Tabuk 71451, Saudi Arabia}
\address[Tata]{Tata Institute of Fundamental Research, Mumbai 400005, India}
% ...
\address[Tokyo]{Department of Physics, University of Tokyo, Tokyo 113-0033, Japan}
% ...

  \begin{abstract}
    We report an improved measurement of $D^0$-$\dbar$ mixing and a search for
    $CP$ violation in $D^0$ decays to $CP$-even final states $K^+K^-$ and $\pi^+\pi^-$.
    The measurement is based on the final Belle data sample of 976~fb$^{-1}$.
    The results are $y_{CP}=(1.11 \pm 0.22 \pm 0.09)\%$ and
    $A_{\Gamma}=(-0.03 \pm 0.20 \pm 0.07)\%$, where the first
    uncertainty is statistical and the second is systematic.
  \end{abstract}


\end{frontmatter}

\section{Introduction}

Mixing of neutral mesons originates from a difference between mass
and flavor eigenstates of the meson-antimeson system. For $D^0$ mesons,
the mass eigenstates are usually expressed as 
$|D^0_{1,2}\rangle = p|D^0\rangle\pm q|\dbar\rangle$
(the sum for $D^0_1$ and the difference for $D^0_2$), with $|p|^2+|q|^2 = 1$. 
The $D^0$-$\dbar$ mixing rate is 
characterized by two parameters: $x = \Delta m/\Gamma$ and $y = \Delta \Gamma/2\Gamma$,
where $\Delta m = m_2 - m_1$ and $\Delta \Gamma = \Gamma_2 - \Gamma_1$
are the differences in mass and decay width,
respectively, between the mass eigenstates $D^0_2$ and $D^0_1$, and $\Gamma$ is the
average $D^0$ decay width. 
If $p = q$, the mass eigenstates are also $CP$ eigenstates;
otherwise, $D^0_{1,2}$ are not \cp\ eigenstates and \cp\ violation
arises in decays of $D^0$ mesons~\cite{charmReview}. 

Mixing in $D^0$ decays to $CP$ eigenstates, such as $D^0 \to K^+K^-$, 
gives rise to an effective lifetime $\tau$ that differs from 
that in decays to flavor eigenstates such as $D^0 \to K^-\pi^+$~\cite{Bergmann}.
The observable
\begin{eqnarray}
  y_{CP} & = & \frac{\tau(D^0 \to K^-\pi^+)}{\tau(D^0 \to K^+K^-)} - 1
\end{eqnarray} 
is equal to the mixing parameter $y$ if \cp\ is conserved\footnote{
  using phase convention $CP|D^0\rangle = -|\dbar \rangle$
}. Otherwise, 
the effective lifetimes of $D^0$ and $\dbar$ decaying to the same $CP$ eigenstate 
differ and the asymmetry
\begin{eqnarray}
  A_{\Gamma} & = & \frac{\tau(\dbar \to K^-K^+)-\tau(D^0 \to K^+K^-)}
  {\tau(\dbar \to K^-K^+)+\tau(D^0 \to K^+K^-)}
\end{eqnarray}
is non-zero. The observables $y_{CP}$ and $A_{\Gamma}$ are, 
in the absence of direct CP violation, 
related to the mixing parameters $x$ and $y$ as~\cite{Bergmann, Nir}
$y_{CP} = \frac{1}{2}(|q/p|+|p/q|)y \cos{\phi} - \frac{1}{2}(|q/p|-|p/q|)x \sin{\phi}$
and
$A_\Gamma = \frac{1}{2}(|q/p|-|p/q|)y \cos{\phi} - \frac{1}{2}(|q/p|+|p/q|)x \sin{\phi}$,
where $\phi = \arg(q/p)$.

The first evidence for $D^0$-$\dbar$ mixing was obtained in 2007 by 
Belle using $D^0 \to K^+K^-$ and $D^0 \to \pi^+\pi^-$~\cite{BelleEvidence} 
and by BaBar using ``wrong-sign'' $D^0 \to K^+\pi^-$ decays~\cite{BabarEvidence}.
These results were later confirmed with high precision by
LHCb~\cite{LHCb} and CDF~\cite{CDF}. The asymmetry
$A_\Gamma$ has been measured by Belle~\cite{BelleEvidence}, BaBar~\cite{BabarAgamma},
CDF~\cite{CDFAgamma} and LHCb~\cite{LHCbAgamma, LHCbAgamma1}.
The measurements of $y_{CP}$ have been reported also
by BaBar~\cite{BabarAgamma}, LHCb~\cite{LHCbycp} and BESIII~\cite{BESIIIycp}.
Here, we report a new measurement of $D^0 \to K^+K^-, \pi^+\pi^-$
decays using almost twice as much data as in Ref.~\cite{BelleEvidence}
and an improved analysis method.
The resolution function now
accounts for a dependence upon polar angle and different
configurations of the silicon vertex detector (see below).

\section{Event selection}

The measurement is based on the final data set of 976~fb$^{-1}$ recorded by the Belle 
detector~\cite{Belle} at the KEKB asymmetric-energy $e^+e^-$~collider~\cite{KEKB}, 
which operated primarily at the center-of-mass energy of
the $\Upsilon(4S)$ resonance, and 60~MeV below. A fraction of the data was recorded 
at the $\Upsilon(1S)$, $\Upsilon(2S)$, $\Upsilon(3S)$, and $\Upsilon(5S)$ resonances; 
these data are included in the measurement.
The Belle detector is described in detail elsewhere~\cite{Belle}.
It includes a silicon vertex detector (SVD), a central drift chamber (CDC),
an array of aerogel Cherenkov counters, and time-of-flight scintillation counters.
Two different SVD configurations were used: a 3-layer configuration for the first 
153~fb$^{-1}$ of data and a 4-layer configuration~\cite{SVD2} for the 
remaining 823~fb$^{-1}$ of data.

The decays $D^0 \to K^+K^-$, $D^0 \to \pi^+\pi^-$ and $D^0 \to K^-\pi^+$ are 
reconstructed in the decay chain $D^{*+} \to D^0\pi^+$, where the charge of the
$D^*$-daughter pion (which has low momentum and thus is referred to as ``slow'')
is used to tag the initial flavor of the $D^0$
meson.\footnote{Throughout this paper, charge-conjugate modes
are included implicitly unless noted otherwise.}
Each final-state charged particle is required to have at least two associated SVD hits 
in each of the longitudinal and azimuthal measuring coordinates. 
To select pion and kaon candidates, we impose
particle identification criteria based on energy deposition in the CDC, the track 
time of flight, and information from the aerogel Cherenkov counters~\cite{PID}.
The identification efficiencies
and the misidentification probabilities are about 85\% and 9\%, respectively, for the
$D^0$ daughters, and about 99\% and 2\%, respectively, for the slow pion 
from $D^{*+}$ decay. The $D^0$ daughters are refitted to a common vertex. 
The $D^0$ production vertex is determined
as the intersection of the $D^0$ trajectory with that of the slow pion,
subject to the constraint that they both originate from the $e^+e^-$
interaction region.
Confidence levels exceeding $10^{-3}$ are required for both fits.
To reject $D$ mesons produced in $B$-meson decays and
also to suppress combinatorial background, the $D^{*+}$ momentum
in the $e^+e^-$ center-of-mass system (CMS) is required to satisfy
$p^*_D > 2.5~{\rm GeV}/c$
for the data taken below the $\Upsilon(5S)$ resonance and 
$p^*_D > 3.1~{\rm GeV}/c$ for the $\Upsilon(5S)$ data.

We select $D^0$ candidates using two kinematic variables: 
the invariant mass $M$ of the $D^0$ and the
energy released in the $D^{*+}$ decay $q=(M_{D^*}-M-m_\pi )c^2$, where
$M_{D^*}$ is the invariant mass of the $D^{*+}$ decay products and
$m_\pi$ is the mass of the charged pion. The proper  decay time of the
$D^0$ candidate is calculated from the projection of the vector
joining the two vertices, $\vec{L}$, onto the $D^0$ momentum vector $\vec{p}$: 
$t=m_{D^0}\vec{L}\cdot\vec{p}/p^2$, where
$m_{D^0}$ is the nominal $D^0$ mass~\cite{PDG}. The proper decay time uncertainty 
$\sigma_t$ of the candidate $D^0$ is evaluated from the error matrices of 
the production and decay vertices.

\begin{figure}[htb]
  \centerline{\includegraphics[width=5cm]{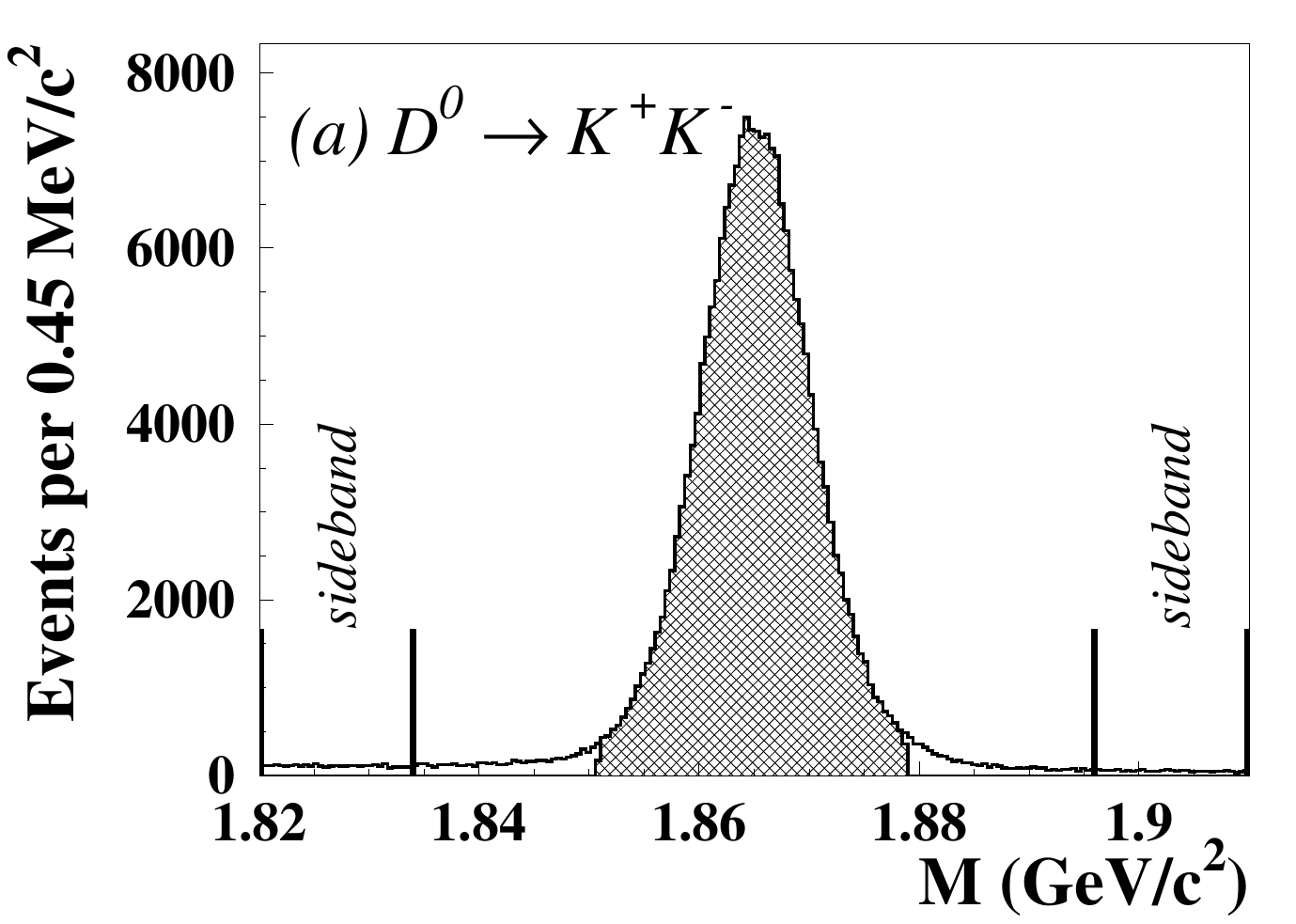}}
  \centerline{\includegraphics[width=5cm]{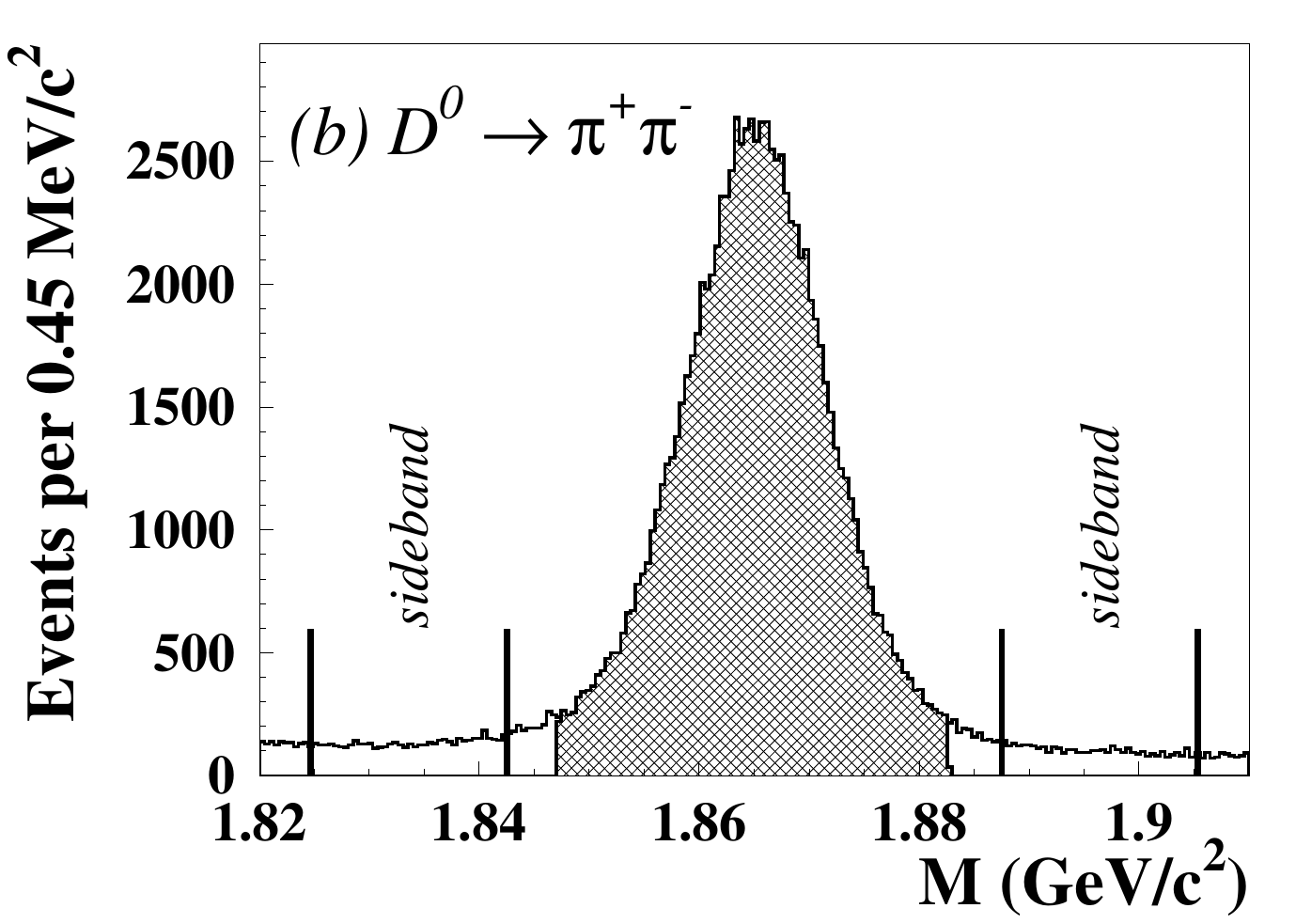}}
  \centerline{\includegraphics[width=5cm]{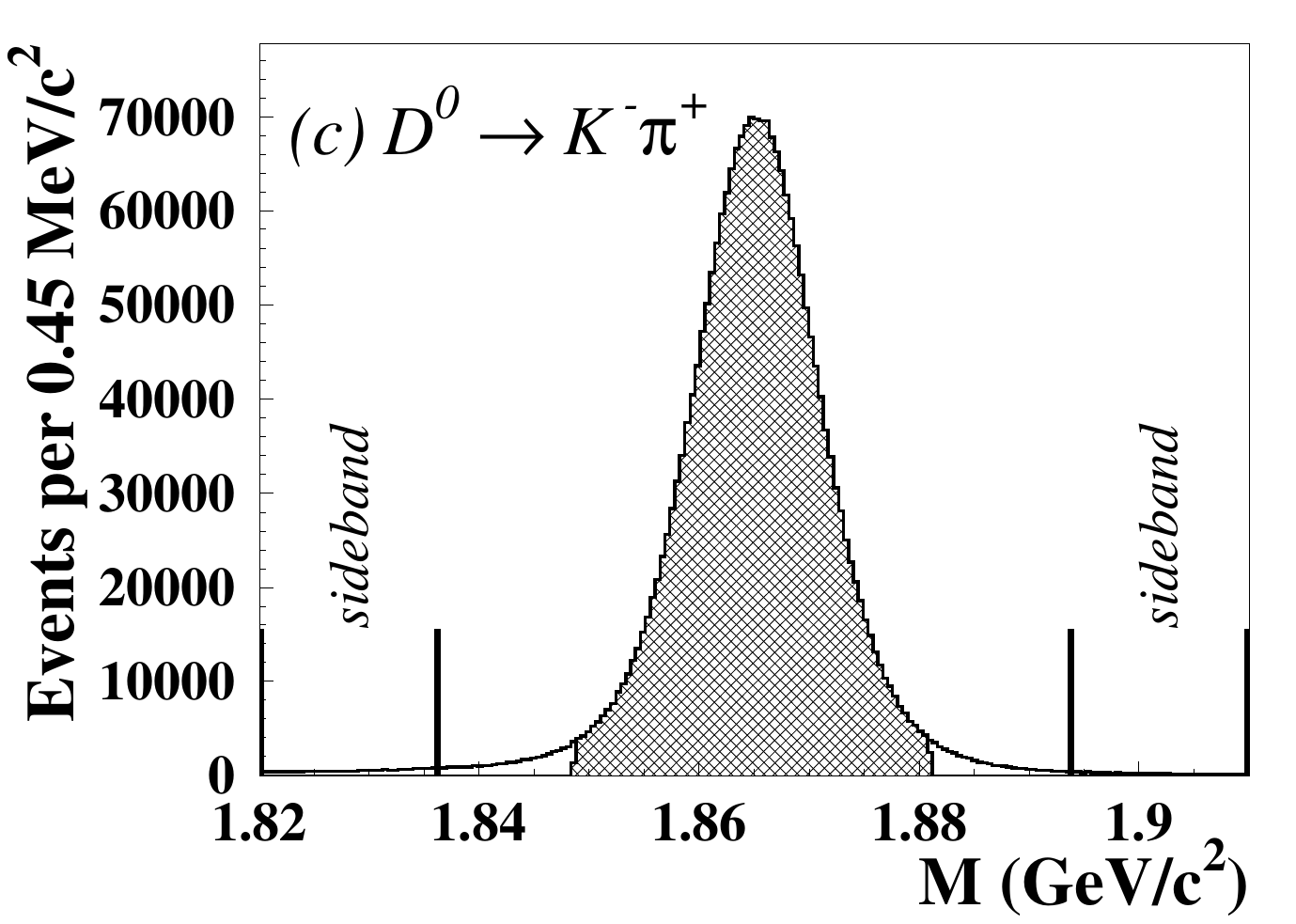}}
  \caption{$D^0$ invariant mass distributions obtained with the 4-layer SVD configuration
    after applying optimized selection criteria on $\Delta q$ and $\sigma_t$. 
    {\it (a)} $D^0 \to K^+K^-$; 
    {\it (b)} $D^0 \to \pi^+\pi^-$; and 
    {\it (c)} $D^0 \to K^-\pi^+$. 
    The shaded regions indicate events selected for the measurement.
    The sideband positions are also indicated.
    \label{massPlots}
  }
\end{figure}

The samples of events for the lifetime measurements are selected
using variables $\Delta M \equiv M - m_{D^0}$, $\Delta q = q - q_0$, and $\sigma_t$, 
where $q_0$ is the nominal energy released in the $D^{*+}$ decay (5.86~MeV).
These selection criteria are optimized using Monte Carlo (MC) simulation
by minimizing the statistical uncertainty on $y_{CP}$.
The simulation is based on EvtGen~\cite{EvtGen} and Pythia generators~\cite{Pythia};
simulated events were processed through a full Belle detector simulation using
Geant~3~\cite{Geant3} and Fluka~\cite{Fluka} to simulate hadronic interactions.
The optimization gives the following selection criteria:
$|\Delta M| < 2.25 \sigma_M$ for all events, where $\sigma_M$ is the r.m.s.\ width
of the $D^0$ invariant mass peak;
$|\Delta q| < 0.66~{\rm MeV}$ and $\sigma_t < 440~{\rm fs}$ 
for the 3-layer SVD configuration; and 
$|\Delta q| < 0.82~{\rm MeV}$ and $\sigma_t < 370~{\rm fs}$ 
for the 4-layer SVD configuration.
The $D^0$ peak, shown in Fig.~\ref{massPlots}, is not purely Gaussian in shape. 
In addition, the width $\sigma_M$ depends on the decay mode and on the
SVD configuration. Typically $\sigma_M \approx 6\!-\!8~{\rm MeV}/c^2$. 

Background is estimated from sidebands in $M$. The sideband position 
is optimized using MC simulation in order to minimize systematic
uncertainties arising from small differences between the 
decay time distribution of events in the sideband and that 
of background events in the signal region.
The sideband windows are shown in Fig.~\ref{massPlots}.
The yields of selected events 
are $242 \times 10^3$ $K^+K^-$, $114 \times 10^3$ $\pi^+\pi^-$,
and $2.61 \times 10^6$ $K^-\pi^+$, with signal purities of
98.0\%, 92.9\% and 99.7\%, respectively. The dominant background is combinatorial.

\section{Lifetime fit}

The measurement is performed by doing a
simultaneous binned maximum likelihood fit to five data samples:
$D^0 \to K^+K^-$, $\dbar \to K^+K^-$,
$D^0 \to \pi^+\pi^-$, $\dbar \to \pi^+\pi^-$, and the sum of
$D^0 \to K^-\pi^+$ and $\dbar \to K^+\pi^-$.
The proper decay time distribution is parameterized as
\begin{eqnarray}
  F(t) & = & \frac{N}{\tau}\int_0^\infty e^{-t'/\tau} R(t-t') dt' + B(t),
  \label{timedistr.eq}
\end{eqnarray}
where $\tau$ is the effective lifetime, $N$ is the signal yield,
$R(t)$ is a resolution function,
and $B(t)$ is the background contribution that is fixed from a fit to the
sideband distribution. 
The decay time acceptance is studied with MC simulations and found
to be constant to good precision within the selected range.

The construction of the resolution function is similar to that of our previous
analysis~\cite{BelleEvidence} but improved to take into account
a possible shape asymmetry
and $D^0$ polar angle dependence. It is constructed 
using a normalized distribution of $\sigma_t$: for each
$\sigma_t$ bin, a common-mean double- or triple-Gaussian probability density function
is constructed.
The fractions $w_k$ and widths $\sigma_k^{\rm pull}$ of these
Gaussian distributions are obtained from fits to the MC distribution
of pulls, defined as $(t-t_{\rm gen})/\sigma_t$, where $t$ and
$t_{\rm gen}$ are the reconstructed and generated proper decay times,
respectively, of simulated $D^0$ decays. 
The resolution function is
\begin{eqnarray}
  R(t) & = & \sum_{i=1}^{n} f_i \sum_{k=1}^{n_g} w_k G(t;\mu_i, \sigma_{ik})\,,
  \label{resofun.eq}
\end{eqnarray}
where $G(t;\mu_i, \sigma_{ik})$ is a Gaussian distribution of mean $\mu_i$
and width $\sigma_{ik}$; 
$f_i$ is the fraction of events in the $i$-th bin of the $\sigma_t$ distribution;
the index $k$ runs over the number of Gaussians $n_g$ used for bin $i$; and
the index $i$ runs over the number of $\sigma_t$ bins.
The means and widths of the Gaussians are parameterized as
\begin{eqnarray}
  \mu_i=t_0 + a (\sigma_i - \overline{\sigma_t}) & \hskip0.30in & 
  \sigma_{ik}= s_k \sigma_k^{\rm pull}\sigma_i\,,
  \label{mean-wid.eq}
\end{eqnarray}
where 
$t_0$ is a resolution function offset, 
$a$ is a parameter to model a possible
asymmetry of the resolution function, 
$\sigma_i$ is the bin central value,
$\overline{\sigma_t}$ is the mean of the $\sigma_t$ distribution,
and
$s_k$ is a width-scaling factor.
The parameters $s_k$, $t_0$ and $a$, in addition to $N$ and $\tau$,
are free parameters in the fit.
To construct $R(t)$ with Eq.~\ref{resofun.eq}, a sideband-subtracted 
$\sigma_t$ distribution is used.

\begin{figure}[t]
  \centerline{\includegraphics[width=6cm]{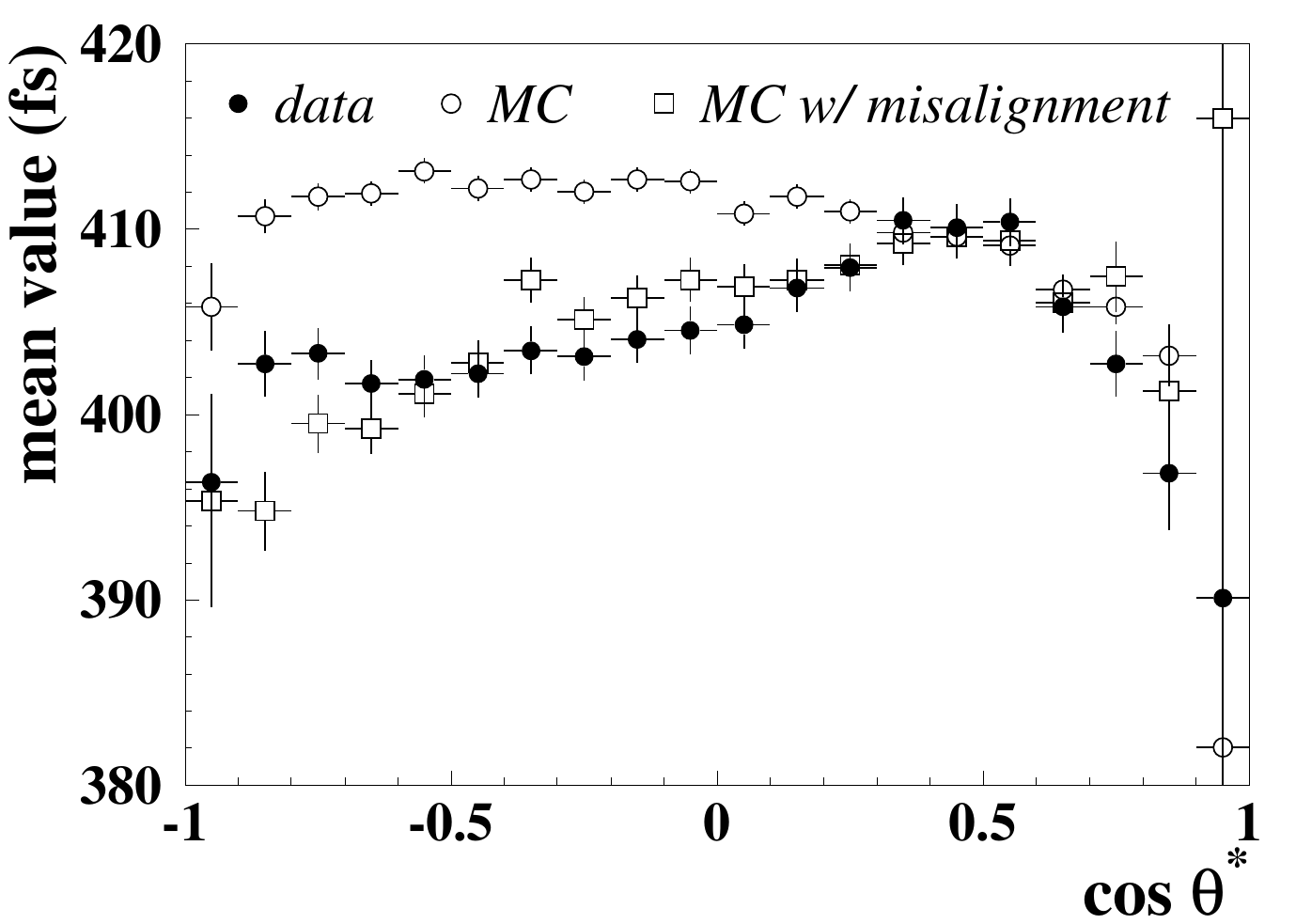}}
  \caption{
    Mean of the sideband-subtracted proper decay time distribution of $D^0 \to K^-\pi^+$ 
    decays as a function of $\cos \theta^*$ for 4-layer SVD data (full circles)
    and corresponding MC simulation (open circles) and for one of the MC samples
    with misaligned SVD (open squares) that shows a dependence similar to data.
    Similar behaviour is observed also for 3-layer SVD configuration.
    \label{meanTime}
  }
\end{figure}

From studies of the proper decay time distribution of
$D^0 \to K^-\pi^+$ decays, we observe a significant dependence
of its mean value on $\cos \theta^*$ (see Fig.~\ref{meanTime}), where $\theta^*$ is the
polar angle of $D^0$ in CMS with respect to the direction of $e^+$.
From MC studies, we find that this effect is due to a small
misalignment of the SVD detector. The effect can be corrected
for when fitting for the lifetime by allowing the resolution function
offset $t_0$ to vary with $\cos \theta^*$.
We thus measure $y_{CP}$ and $A_\Gamma$ in bins of $\cos \theta^*$, with
the resolution function calculated separately for each bin.
An additional requirement $|\cos \theta^*|<0.9$ is imposed to
suppress events with large offsets (about 1\% of events).

The background term in Eq.~\ref{timedistr.eq} is parameterized as the
sum of a component with zero lifetime and
a component with an effective lifetime $\tau_b$:
\begin{equation}
  B(t)\ =\ N_b \int_0^\infty [p \delta(t') + (1-p) \frac{1}{\tau_b}e^{-t'/\tau_b}]R_b(t-t') dt'\,.
\end{equation}
The resolution function $R_b(t)$ is also parameterized with
Eq.~\ref{resofun.eq} except that, for each $\sigma^{}_t$ bin, the function
is taken to be symmetric ($a = 0$) and always composed of three Gaussians, 
with the second and third scaling factors being equal ($s_2 = s_3$). 
The $\sigma_t$ distribution is taken from an $M$ sideband.
The fraction $p$ of the zero-lifetime component is
found to be $\cos \theta^*$-dependent; its value is
fixed in each bin using MC simulation. 
The parameters $t_0$, $s_1$, $s_2$ and $\tau_b$ are determined 
separately for each decay mode and SVD configuration
from a fit to sideband distributions summed over $\cos \theta^*$ bins. 
However, the background shape 
is still $\cos \theta^*$ dependent, because the $\sigma_t$ distribution,
the zero-lifetime fraction $p$ and the yield $N_b$ all
depend on $\cos \theta^*$. The quality of these fits exceeds 15\% confidence level (CL).

To extract $y_{CP}$ and $A_\Gamma$, the decay modes are fitted simultaneously in
each $\cos \theta^*$ bin and separately for each of the two SVD configurations.
The parameters shared among the decay modes are $y_{CP}$ and
$A_\Gamma$ (between $KK$ and $\pi\pi$), $t_0$ and $a$ (among
all decay modes), and parameters $s_1$, $s_2$ and $s_3$, up to
an overall scaling factor. Results for individual $\cos\theta^*$
bins and for the two data sets are combined into an overall result via
a least-squares fit to a constant. 

The fitting procedure is tested with a generic MC sample
equivalent to six times the data statistics. The fitted $y_{CP}$
and $A_\Gamma$ are consistent with the input zero
value, and the fitted $K\pi$ lifetime is consistent with
the generated value. 
Linearity tests performed with MC-simulated events re-weighted to reflect
different $y_{CP}$ and $A_\Gamma$ values show no bias.

\begin{figure}[htb]
  \centerline{\includegraphics[width=4.3cm]{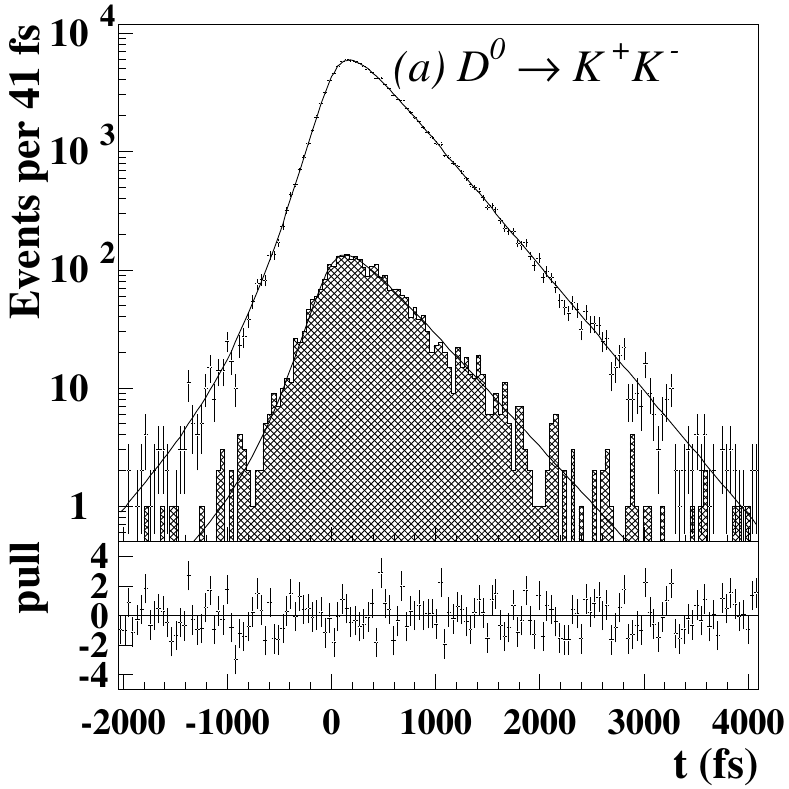}
    \includegraphics[width=4.3cm]{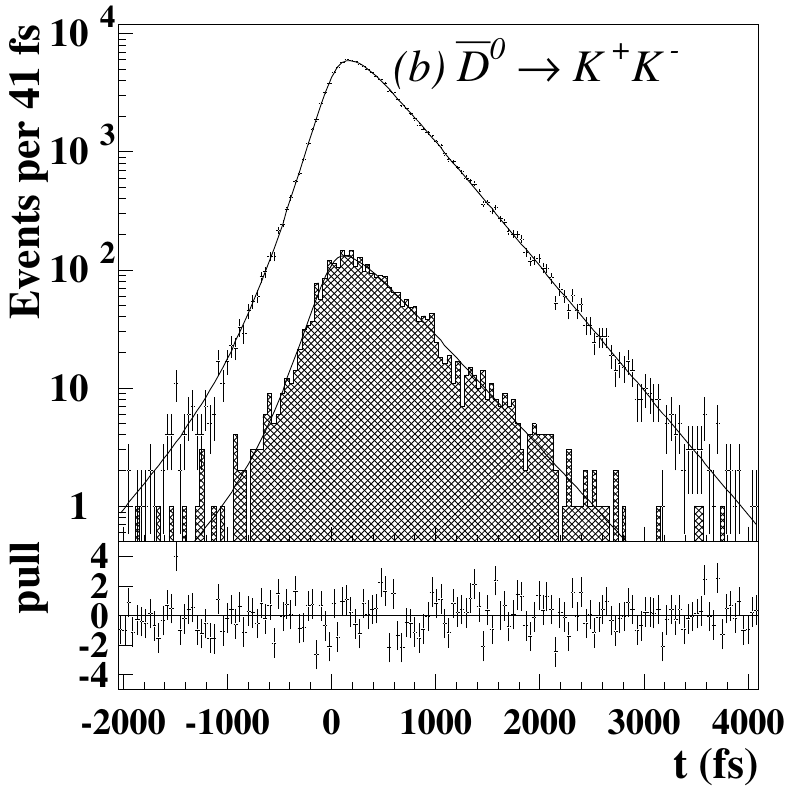}}
  \centerline{\includegraphics[width=4.3cm]{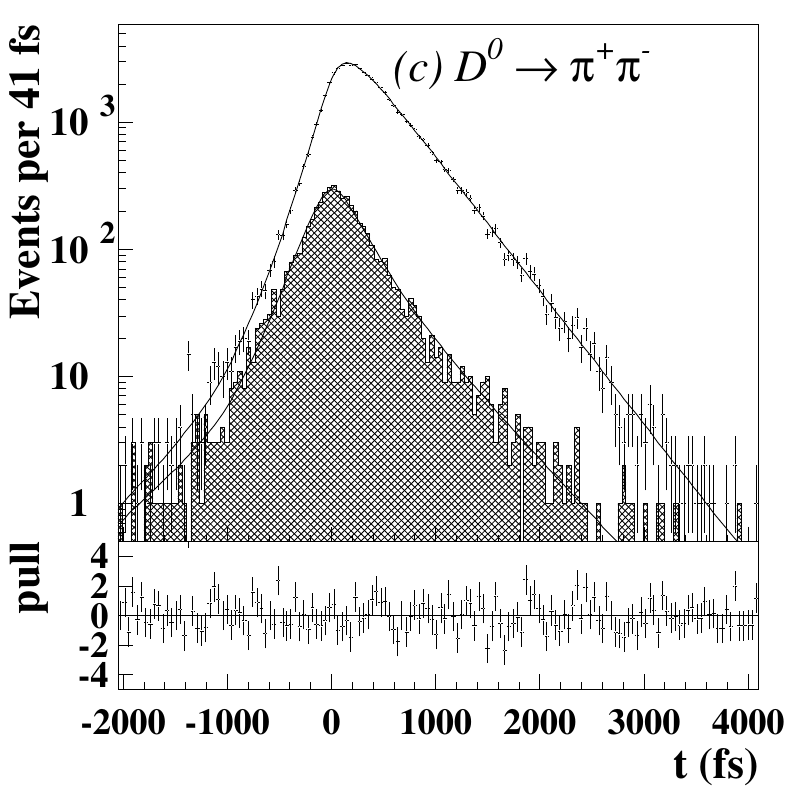}
    \includegraphics[width=4.3cm]{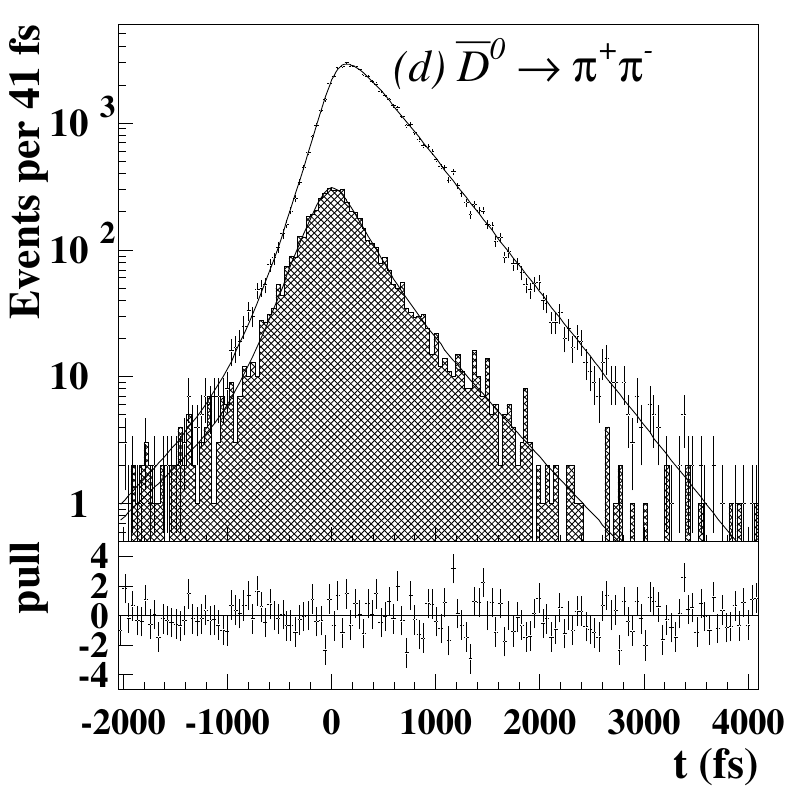}}
  \centerline{\includegraphics[width=4.3cm]{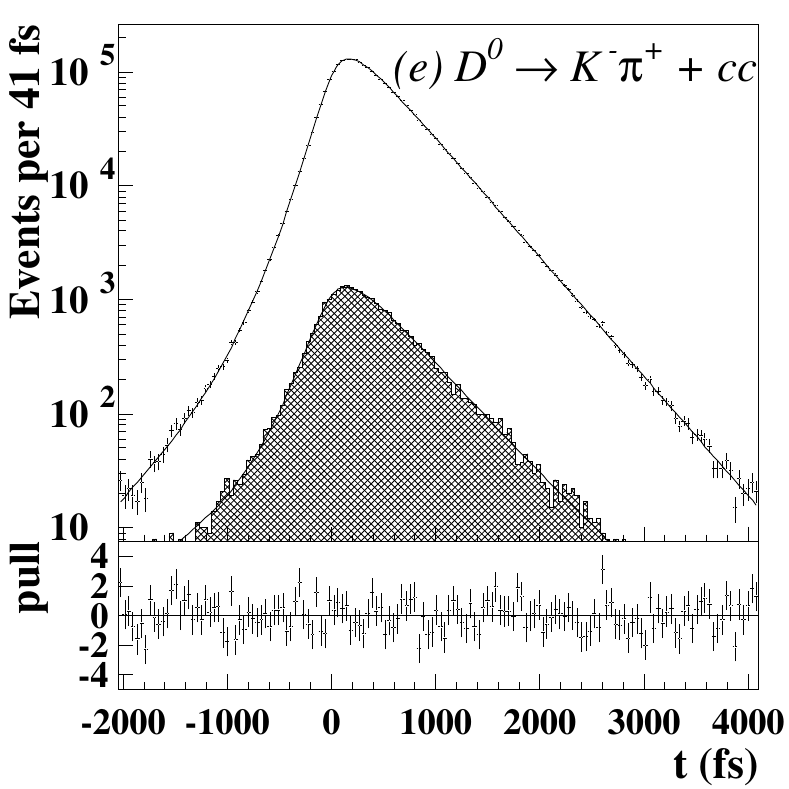}}
  \caption{
    Proper decay time distributions summed over $\cos \theta^*$ bins and 
    both running periods with the sum of fitted functions superimposed.
    Shown as error bars are the distributions of events in the $M$ signal region
    while the shaded area represents background contributions
    as obtained from $M$ sidebands.
    The plots beneath the distributions show the
    pulls of simultaneous fit (i.e., residuals divided by errors).
    \label{timeFits}
  }
\end{figure}

The fitting procedure is then applied to the measured data.
The fitted proper decay time distributions summed over $\cos \theta^*$
bins and running periods with the two SVD configurations 
are shown in Fig.~\ref{timeFits}. 
The pulls, plotted beneath each fitted distribution, show no significant structure. 
The normalized $\chi^2$ is 1.13.\footnote{We use Pearson's definition of 
$\chi^2$ and take only the bins with the fitted function greater than one.}
The confidence levels of individual fits in bins of $\cos \theta^*$ are 
above 5\%, except for one with CL=3.3\%, and are distributed uniformly. 

\begin{figure}[htb]
  \centerline{\includegraphics[width=6cm]{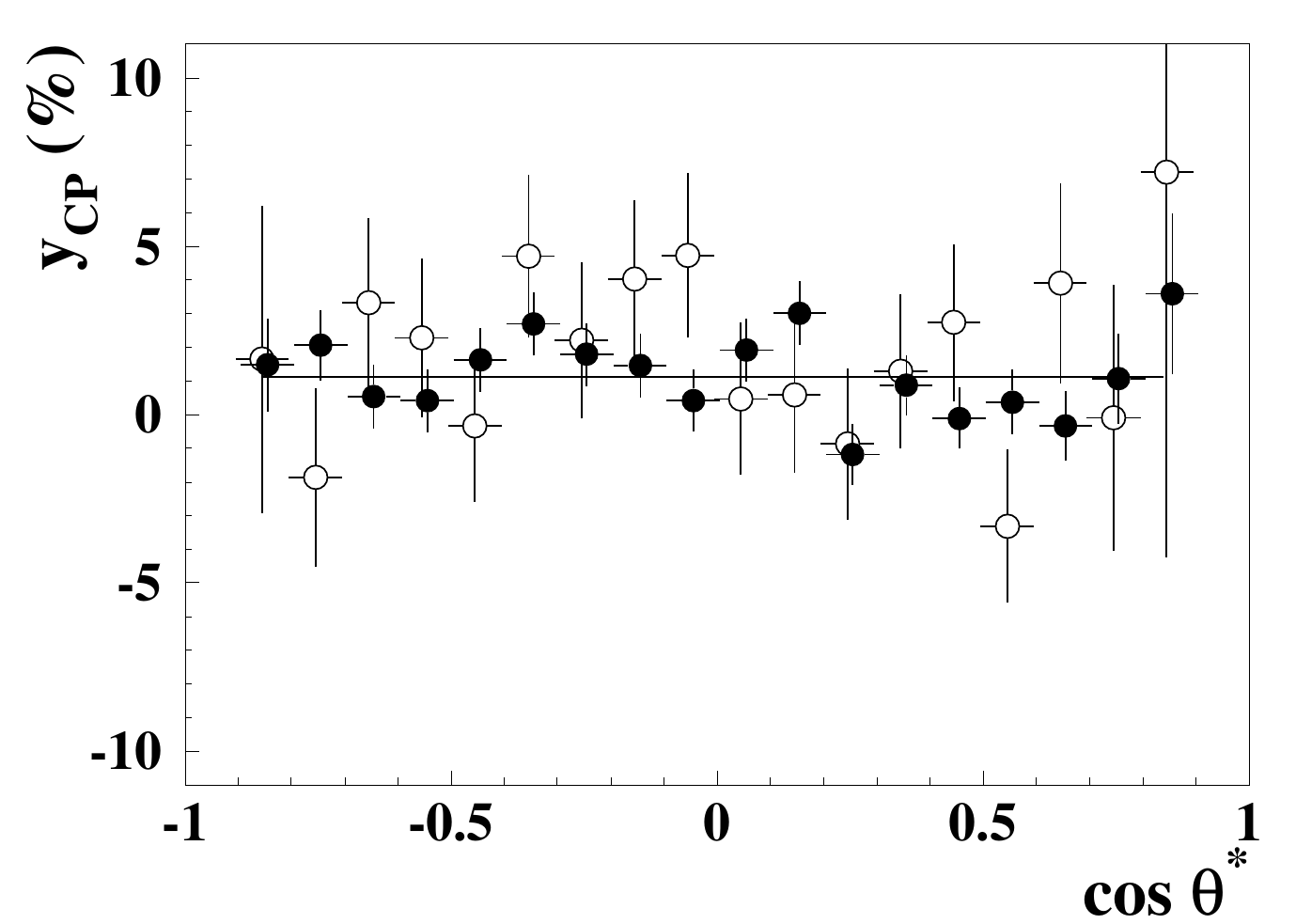}}
  \caption{Fitted $y_{CP}$ in bins of $\cos \theta^*$ for 3-layer SVD data (open circles)
    and for 4-layer SVD data (full circles). 
    The horizontal line is the result of fitting the points to a constant.
    \label{ycpFit}
  }
\end{figure}

\begin{figure}[htb]
  \centerline{\includegraphics[width=6cm]{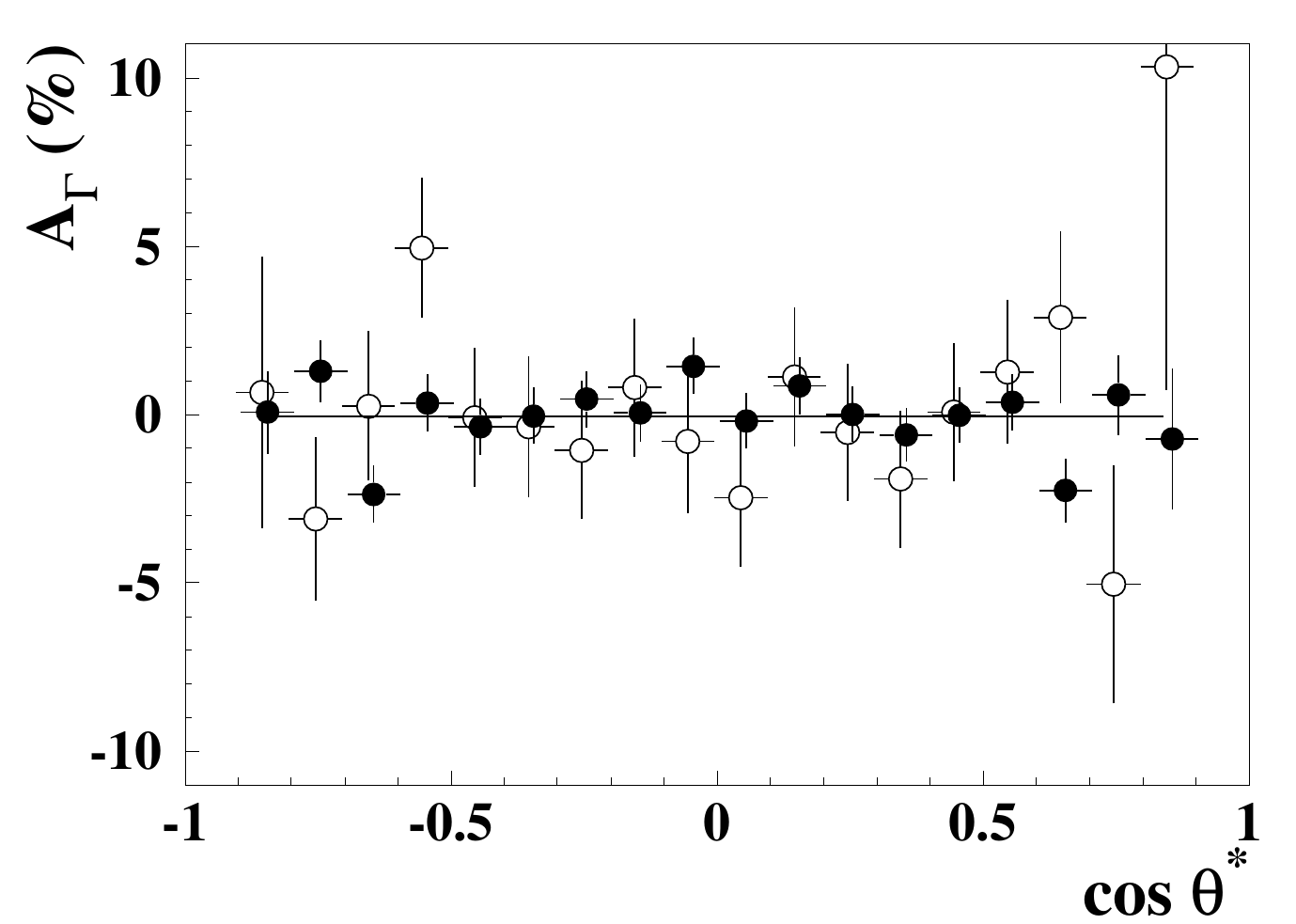}}
  \caption{Fitted $A_\Gamma$ in bins of $\cos \theta^*$ for 3-layer SVD data 
    (open circles)
    and for 4-layer SVD data (full circles).
    The horizontal line is the result of fitting the points to a constant.
    \label{agammaFit}
  }
\end{figure}

The fitted values of $y_{CP}$ and $A_\Gamma$ in bins of $\cos \theta^*$
are shown in Figs.~\ref{ycpFit} and~\ref{agammaFit}. The values obtained
with a least-squares fit to a constant are $y_{CP} = (1.11\pm 0.22)\%$ and
$A_\Gamma = (-0.03 \pm 0.20)\%$, where the uncertainties are statistical only; the
confidence levels are 32\% and 40\%, respectively. The fitted $D^0$ lifetime
is $(408.46\pm 0.54)~{\rm fs}$ (statistical uncertainty only), which is consistent
with the current world average of $(410.1\pm 1.5)$~fs~\cite{PDG}. 

\section{Systematic uncertainties}

The estimated systematic uncertainties are listed in Table~\ref{syst.tab}.
The impact of imperfect SVD alignment is studied with a dedicated
signal MC simulation in which different local and global SVD
misalignments are modeled.
Local misalignment refers to a random translation and rotation of 
each individual silicon strip detector according to the alignment precision, 
while global misalignment refers to a translation 
and rotation of the entire SVD with respect to the CDC.
The systematic uncertainties are taken to be the r.m.s. of the differences 
between these results and the nominal result that assumes perfect SVD alignment.
We obtain 0.060\% for $y_{CP}$ and 0.041\% for $A_\Gamma$. 

The uncertainty due to the position of the mass window is 
estimated by varying the position of the window by the
small differences found between MC simulation and data 
in the position of the $D^0$ mass peak, about $\pm1~{\rm MeV}/c^2$.
This resulting uncertainty is relatively small:
0.007\% for $y_{CP}$ and 0.009\% for $A_\Gamma$. 

Background contributes to the systematic uncertainty in two ways:
statistical fluctuations of sideband distributions and modeling. 
The former is found to contribute 0.051\% for $y_{CP}$ and 0.050\%
for $A_\Gamma$. The latter arises from modeling the background
distribution with that of 
sideband events; this uncertainty is estimated from MC simulation to
be 0.029\% for $y_{CP}$ and 0.007\% for $A_\Gamma$. Combining the
two contributions in quadrature gives total uncertainties of 0.059\%
for $y_{CP}$ and 0.050\% for $A_\Gamma$.

Systematics due to the resolution function are estimated 
using two alternative parameterizations in the fit:
one in which the parameter $a$ in Eq.~\ref{mean-wid.eq} 
is fixed to zero, and the other in which this parameter
is floated but not shared among different decay modes. 
We find variations of 0.030\% for $y_{CP}$ and 0.002\% for $A_\Gamma$.
Systematics due to binning are estimated by varying the number of bins
in $\cos \theta^*$ and $t$. This contribution is found to be 0.021\% for
$y_{CP}$ and 0.010\% for $A_\Gamma$. 

Possible acceptance variations with decay time are tested by fitting 
decay time distributions of MC events that pass the selection criteria. 
We always recover the generated lifetimes, for all decay
modes, indicating uniform acceptance.
We conclude that this effect is negligible.
All individual contributions are
added in quadrature to obtain overall systematic uncertainties 
of 0.09\% for $y_{CP}$ and 0.07\% for $A_\Gamma$.

\begin{table}[htb]
  \caption{Systematic uncertainties.}
  \label{syst.tab}
  \begin{center}
    \renewcommand{\arraystretch}{1.2}
    \begin{tabular}
      {@{\extracolsep{\fill}} l | c c }
      \hline \hline
      Source & $\Delta y_{CP}$ (\%) & $\Delta A_\Gamma$ (\%) \\
      \hline
      SVD misalignment     & 0.060 & 0.041 \\
      Mass window position & 0.007 & 0.009 \\
      Background           & 0.059 & 0.050 \\
      Resolution function  & 0.030 & 0.002 \\
      Binning              & 0.021 & 0.010 \\
      \hline
      Total                & 0.092 & 0.066 \\
      \hline\hline
    \end{tabular}
  \end{center}
\end{table}

\section{Conclusions}

Using the final Belle data set, we measure the difference
from unity of the ratio of lifetimes 
of $D^0$ mesons decaying to $CP$-even eigenstates 
$K^+ K^-,\,\pi^+\pi^-$ and to the flavor eigenstate $K^-\pi^+$.
Our result is
\begin{eqnarray}
  y_{CP} & = & [+1.11 \pm 0.22 {\rm\ (stat.)} \pm 0.09 {\rm\ (syst.)}]\%\,.
\end{eqnarray}
The significance of this measurement is 4.7$\sigma$ when both
statistical and systematic uncertainties are combined in quadrature.
We also search for \cp\ violation, measuring a \cp\ asymmetry
\begin{eqnarray}
  A_\Gamma & = & [-0.03 \pm 0.20 {\rm\ (stat.)} \pm 0.07 {\rm\ (syst.)}]\%\,.
\end{eqnarray}
This value is consistent with zero. These results are significantly
more precise than our previous results~\cite{BelleEvidence} and supersede them.
They are compatible with results from other experiments~\cite{BabarAgamma,
CDFAgamma,  LHCbAgamma, LHCbAgamma1, LHCbycp, BESIIIycp}
and the world average values~\cite{HFAG}.

\section*{Acknowledgments}

We thank the KEKB group for the excellent operation of the
accelerator; the KEK cryogenics group for the efficient
operation of the solenoid; and the KEK computer group,
the National Institute of Informatics, and the 
PNNL/EMSL computing group for valuable computing
and SINET4 network support.  We acknowledge support from
the Ministry of Education, Culture, Sports, Science, and
Technology (MEXT) of Japan, the Japan Society for the 
Promotion of Science (JSPS), and the Tau-Lepton Physics 
Research Center of Nagoya University; 
the Australian Research Council and the Australian 
Department of Industry, Innovation, Science and Research;
Austrian Science Fund under Grant No.~P 22742-N16 and P 26794-N20;
the National Natural Science Foundation of China under Contracts 
No.~10575109, No.~10775142, No.~10875115, No.~11175187, and  No.~11475187; 
the Ministry of Education, Youth and Sports of the Czech
Republic under Contract No.~LG14034;
the Carl Zeiss Foundation, the Deutsche Forschungsgemeinschaft
and the VolkswagenStiftung;
the Department of Science and Technology of India; 
the Istituto Nazionale di Fisica Nucleare of Italy; 
National Research Foundation (NRF) of Korea Grants
No.~2011-0029457, No.~2012-0008143, No.~2012R1A1A2008330, 
No.~2013R1A1A3007772, No.~2014R1A2A2A01005286, No.~2014R1A2A2A01002734, 
No.~2014R1A1A2006456;
the Basic Research Lab program under NRF Grant No.~KRF-2011-0020333, 
No.~KRF-2011-0021196, Center for Korean J-PARC Users, No.~NRF-2013K1A3A7A06056592; 
the Brain Korea 21-Plus program and the Global Science Experimental Data 
Hub Center of the Korea Institute of Science and Technology Information;
the Polish Ministry of Science and Higher Education and 
the National Science Center;
the Ministry of Education and Science of the Russian Federation and
the Russian Foundation for Basic Research;
the Slovenian Research Agency;
the Basque Foundation for Science (IKERBASQUE) and 
the Euskal Herriko Unibertsitatea (UPV/EHU) under program UFI 11/55 (Spain);
the Swiss National Science Foundation; the National Science Council
and the Ministry of Education of Taiwan; and the U.S.\
Department of Energy and the National Science Foundation.
This work is supported by a Grant-in-Aid from MEXT for 
Science Research in a Priority Area (``New Development of 
Flavor Physics'') and from JSPS for Creative Scientific 
Research (``Evolution of Tau-lepton Physics'').


\bibliographystyle{elsarticle-num}

\end{document}